\documentclass[english,twoside,a4paper,10pt]{article}

\usepackage[latin1]{inputenc}
\usepackage[T1]{fontenc}
\usepackage{amsmath}
\usepackage{amsfonts}
\usepackage{graphicx}
\usepackage{subfigure}
\usepackage{a4wide}
\usepackage{amssymb}
\usepackage{fancyhdr}
\usepackage{mathrsfs}
\begin{document}

\title{\bf Looking for empty topological wormhole spacetimes in $F(R)$-modified gravity }
\author{
R. Di Criscienzo\footnote{E-mail address: roberto.dicriscienzo@gmail.com},
R. Myrzakulov$^{1, }$\footnote{E-mail address: myrzakulov@gmail.com}
and
L.~Sebastiani$^{1, }$\footnote{E-mail address: l.sebastiani@science.unitn.it}
\\
\begin{small}
$^1$ Eurasian International Center for Theoretical Physics and  Department of General
\end{small}\\
\begin{small} 
Theoretical Physics, Eurasian National University, Astana 010008, Kazakhstan
\end{small}\\
}

\date{}

\maketitle

%\def\thesection{\Roman{section}}
%\def\theequation{\Roman{section}.\arabic{equation}}

%%%%%%%%%%%%%%%%%%%%%
%  Abstract
%%%%%%%%%%%%%%%%%%%%%
\begin{abstract}
\noindent Much attention has been recently devoted to modified theories of gravity
%in the attempt to efficiently describe both early inflation and late-time acceleration of our universe without referring to the cosmological constant or other {\it ad hoc} kinds of fluids. 
 the simplest models of which overcome General Relativity simply by replacing $R$ with $F(R)$ in the Einstein--Hilbert action. Unfortunately, such models typically lack most of the beautiful solutions discovered in Einstein's gravity. Nonetheless, in $F(R)$ gravity,  it has been possible to get at least few black holes, but still we do not know any empty wormhole-like spacetime solution. The present paper aims to explain why it is so hard to get such solutions (given that they exist!). Few solutions are derived in the simplest cases while only an implicit form has been obtained in the non-trivial case. 
\end{abstract}
%%%%%%%%%%%%%%%%%%%%%

%----------------------------
%PACS
%----------------------------

%\def\thesection{\Roman{section}}
%\def\theequation{\Roman{section}.\arabic{equation}}
%===========================================================================

\section{Introduction}

In the last years, much attention has been paid to the so-called modified gravity 
theories in the attempt to unify the early-time inflation 
with the late-time acceleration of the universe (for recent review, see 
~\cite{review}). The simplest class of such theories is given by $F(R)$-gravity, where the action is given by a general function $F(R)$ of the the Ricci scalar $R$. If some modified theory lies behind our universe, it may be interesting to explore its mathematical structure and the possibility to recover features and solutions of General Relativity in its framework.  It is in this sense that, for example, we have to interpret current investigations of black holes in modified theories of gravity \cite{bellini}. Among the amenities of General Relativity there are topologically non-trivial spacetimes like wormholes (cf. Ref. \cite{visser} for an exhaustive introduction and Ref.~\cite{B} and references therein for recent works). One can imagine a wormhole as a three-dimensional space with two spherical holes in it, connected one to another, by means of a ``handle'' \cite{frolov}. What likes of such objects is the possibility of entering the wormhole and exiting into external space again: a property which sensibly distinguishes (traversable) wormholes from black holes.  Interest in this field arose twenty-five years ago or so, as it was found that stable wormholes could be transformed into time machines \cite{mt}. It is worth to note that even the Schwarzschild metric with an appropriate choice of topology, describes a wormhole, but not a traversable one. In fact, in order to prevent the wormhole's throat to pinch off so quickly that it cannot be traversed in even one direction, it is necessary to fill the wormhole with non-zero stress and energy of unusual nature. By ``unusual'', we mean here that a stable, traversable wormhole is supported against (attractive) gravity by matter which violates some energy condition (typically, the weak energy condition)... or better, this is what occurs in General Relativity! In $F(R)$-theories of gravity new scenarios are possible: as shown in \cite{Jam}, it is always possible to get a wormhole with matter preserving at least one energy condition (e.g. the strong energy condition) just choosing in an opportune way the metric components. If this procedure may look quite artificial, then following  \cite{W,Furey:2004rq} it is possible to show that models as $F(R)=\sum_n a_n R^n$ for suitable integer $n$ display a matter behavior close to the wormhole throat such to respect the energy conditions and to prevent large anisotropies -- typical features of Einstein's wormholes. The existence of necessary conditions for having wormholes which respect the weak energy condition and possibly the strong energy condition has been studied in \cite{bip} at least in polynomial models of the third order or higher. 
The interplay between $F(R)$-gravity and scalar-tensor theories has been studied in \cite{Staro} with respect to the case of wormhole solutions. 
Lobo and Oliveira construct in \cite{cond} traversable wormhole geometries in the context of $F(R)$ modified theories of gravity imposing that the matter threading the wormhole satisfies the energy conditions, so that it is the effective stress-energy tensor containing higher order curvature derivatives that is responsible for the null energy condition violation. In particular, by considering specific shape functions and several equations of state, exact solutions for $F(R)$ are found.

However, what still lacks in the physics of $F(R)$ wormholes is an empty solution where the role of repulsive gravity is played by geometry and no matter is necessary to support the solution. 
As we shall see, giving an explicit empty wormhole solution is all but easy.  Few new solutions will be found in the case $R=0$; the highly non-trivial case being still inaccessible with respect to analytic techniques. Still we are confident that, also thanks to this work, numeric solutions are at disposal in the near future.

The organization of the paper is as follows. In Section {\bf 2} we will derive the field equations of topological static spherical symmetric solutions in $F(R)$-gravity by using a method based on Lagrangian multipliers which permits to deal with a system of ordinary equations and we will see how is possible to reconstruct the models by starting from the solutions. Important classes of $F(R)$-black hole solutions can be found in this way. In Section {\bf 3} the structure of the metric able to realize traversable wormholes is introduced in two equivalent forms. In Section {\bf 4}, effective energy conditions are investigated and it is shown that the equivalent description of modified gravity as an effective fluid violates the weak energy condition on the throat. In Section {\bf 5}  some wormhole solutions in empty space are found in the framework of $F(R)$-modified gravity. These solutions are characterized by null or constant Ricci scalar and can be realized by a large class of models. In Section {\bf 6}, solutions with non zero curvature are discussed and the implicit form of $F(R)$-models which realizes these solutions is derived. Final remarks are given in Section {\bf 7}.

%%%%%%%%%%%%%%%%%%%%%%%%%%%%%%%%%%%%%%%%%%%%%%%%%%%%%%%%%%%%%%%%%%%%%%%%%%%%%%%%%%%%%%%%%%%%%%%%%%%%%%%%%%%%%%%%%%%%%%%%%%%%%%%%%%%
We use units where $k_{B}=c=\hbar=1$ and denote the gravitational constant
$\kappa^2=8\pi G_N\equiv8\pi/M_{Pl}^2$ with the Planck mass of
$M_{PL}=G^{-1/2}_N=1.2\times 10^{19}\text{GeV}$.
%%%%%%%%%%%%%%%%%%%%%%%%%%%%%%%%%%%%%%%%%%%%%%%%%%%%%%%%%%%%%%%%%%%%%%%%%%%%%%%%%%%%%%%%%%%%%%%%%%%%%%%%%%%%

\section{ Topological SSS vacuum solutions in $F(R)$-gravity}

In this Section we derive the equations of motion (EoMs)
for topological static spherical symmetric (SSS) solutions in $F(R)$-gravity.
We will write the metric in a general form in order to use it to investigate vacuum wormholes. 
To derive our equations we use a method based on Lagrangian multipliers, which permits to deal with
a system of ordinary differential equations (see Ref. \cite{Lmulti} and Ref. \cite{Lmulti2} for its application in FRW case).

The action of modified $F(R)$-gravity in vacuum reads 
\begin{equation}
I=\frac{1}{2\kappa^2}\int_{\mathcal{M}} d^4 x\sqrt{-g}F(R)\,,\label{action} 
\end{equation}
where $g$ is the determinant of the metric tensor, $g_{\mu\nu}$, $\mathcal{M}$ is the space-time manifold and $F(R)$ is a generic function of the Ricci scalar $R$.

Let the metric assume the most general  static spherically symmetric topological form,
\begin{equation}
ds^2 = -V(r) dt^2 +\frac{ dr^2}{ B(r)} + r^2 d\sigma_k^2,\label{M1}
\end{equation}
where $V(r)$ and $B(r)$ are functions of $r >0$ only and 
\begin{equation}
d \sigma_k^2:= \frac{d\varrho^2}{1-k\varrho^2} + \varrho^2 d\varphi^2\;, \qquad \varphi \in [0,2\pi) \quad \mbox{and} \quad k=0,\pm 1
\end{equation}
 represents the metric of a topological two-dimensional surface parametrized by $k$, such that the manifold will be either a sphere $S_2$ ($k=1$), a torus $T_2$ ($k=0$) or a compact hyperbolic manifold $Y_2$ ($k=-1$). 
With this {\it ansatz}, the scalar curvature reads
\begin{eqnarray}
-R  &=&\frac{V''(r)}{V(r)}B(r)-\frac{B(r)}{2}\left(\frac{V'(r)}{V(r)}\right)^2+\frac{B'(r)}{2}\frac{V'(r)}{V(r)}+\frac{2 B(r)}{r}\frac{V'(r)}{V(r)}
\nonumber\\
&&+\frac{2B'(r)}{r}-\frac{2k}{r^2}+\frac{2B(r)}{r^2}\,.
\label{R}
\end{eqnarray}
After integration over the transverse metric $d\sigma_k^2$, introducing the Lagrangian multiplier $\lambda$ and thanks to Eq.~(\ref{R}), the  action may be re-written as 
\begin{eqnarray}
\label{tilde action}
& &\hspace{-5mm}\tilde I = \frac{1}{2\kappa^2}\int dt\int d{ r}\,\sqrt{\frac{V(r)}{B(r)}}r^2\,\left\{ F(R)-\lambda \left[R+
\frac{V''(r)}{V(r)}B(r)-\frac{2k}{r^2}+\frac{2B(r)}{r^2} \right.\right.\nonumber\\
& & \left.\left. - \frac{1}{2}B(r)\left(\frac{V'(r)}{V(r)}\right)^2+B'(r)\frac{V'(r)}{2V(r)}+\frac{2 B(r)}{r}\frac{V'(r)}{V(r)}+\frac{2B'(r)}{r}\right]\right\}\,.
\end{eqnarray}
Making the variation with respect to $R$, one gets
\begin{equation}
\lambda=\frac{d}{d R}F(R) \equiv F_R(R)\,.
\label{lambda}
\end{equation}
Substituting Eq. (\ref{lambda}) in (\ref{tilde action}) and making  a partial integration, the effective Lagrangian $\mathscr L$ of the system assumes the form

%\\\phantom{line}\\
\begin{eqnarray}
\mathscr L(r,V,V',B,B',R,R')&=&\sqrt{\frac{V(r)}{B(r)}}\left[\phantom{\frac{0}{0}}(F-F_R R)r^2 +2 F_R\left(k-B(r)-B'(r)r\right)+\right.\nonumber\\
&&\hspace{1.3cm}\left.
+F_R'\left(\frac{V'(r)}{V(r)}\right)B(r)r^2
\right]\,,
\end{eqnarray}
%\phantom{line}
where the prime index $^{\prime}$ denotes derivative with respect to $r$. 
Therefore, the equations of motion are 

%dividing to $V(r)/(2B(r)^2)\sqrt{B(r)/V(r)}$
\begin{equation}
\label{uno0}
\left\{\begin{array}{ll}
0=&r^2(F-F_R R)+2F_R(k-r B'(r)-B(r))-F'_R(r^2B'(r)+4rB(r))-2r^2B(r)F_R''\\ 
\phantom{line} \\
0=&r^2(F-F_R R)+2F_R\left(k-B(r)-B(r)r\frac{V'(r)}{V(r)}\right)- r B(r)F'_R\left(4+r\frac{V'(r)}{V(r)}\right)\,.
\end{array}\right.
\phantom{line}
\end{equation}

The EoMs with Eq.~(\ref{R}) form a system of three ordinary differential equations in the three unknown quantities $V(r)$, $B(r)$ and $R(r)$.

Starting from these equations, we may try to reconstruct SSS solutions realized in $F(R)$--gravity. 
In particular, combining equations (\ref{uno0}) and taking the derivative of the first equation in (\ref{uno0}) with respect to $r$, we end with
\begin{equation}
\label{Seba1}
\left\{\begin{array}{ll}
&F'_R=-\frac{2F_R}{r}+2F_R''\left(\frac{V'(r)}{V(r)}-\frac{B'(r)}{B(r)}\right)^{-1}\\
\\
&0= F'_R\left[  \frac{4B(r)}{r^2}-B''(r)-\frac{4B'(r)}{r}+\frac{V''(r)}{V(r)}B(r) +\frac{V'(r)}{V(r)}\left(\frac{V'(r)}{2V(r)}B(r)+\frac{1}{2}B'(r)+\frac{2B(r)}{r}\right)\right] +\\
\\
&\hspace{1cm}+ F'\left(\frac{4B(r)}{r^3}-\frac{2B''(r)}{r}-\frac{4k}{r^3}\right)-\left(\frac{4B(r)}{r}+3B'(r)\right)F''_R -2B(r)F_R'''\,.
\end{array}\right.
\end{equation}
In this way, we deal with equations that depend on the model only through $F_R$. Of course, one might replace $F_R$ by $F'/R'$, but to the price of dealing with much more involved equations.
In principle, given the model, that is $F_R$ as a function of $r$, equations (\ref{Seba1}) let us to derive both $V(r)$ and $B(r)$, {\it i.e.} the explicit form of the metric. 

A remark is still in order about Eq. (\ref{Seba1}): also if we have assumed from the beginning that $V(r)\neq B(r)$, in case $V(r)=B(r)$ the equation becomes $F''_R=0$ recovering the important class of black hole solutions already discussed in Ref.~\cite{Lmulti} and in Ref.~\cite{Iran}. 

Let us see how the reconstruction method works with a simple class of solutions.
We can consider for example $F_R(r)\propto (r/\tilde r)^q$, $q$ being a fixed parameter and $\tilde r$ a dimensional constant. From Eq. (\ref{Seba1}) one has 
\begin{equation}
\frac{V'(r)}{V(r)}=\frac{2q(q-1)}{2+q}\frac{1}{r} +\frac{B'(r)}{B(r)}\,.\label{VL}
\end{equation}
By taking into account that 
$$\frac{V''(r)}{V(r)}=\frac{d}{dr } \left( \frac{V'(r)}{V(r)}\right)+\left(\frac{V'(r)}{V(r)}\right)^2\,,$$
 we can solve Eq. (\ref{Seba1}) and obtain ($C_0, C_1$ being constants of integration)
\begin{equation}
B(r)=r^{a_1}(C_0+C_1 r^{a_2})-\frac{k(2+q)^2}{2(q^4-q^3-3q^2-4q-2)}\,,
\end{equation}
where 
\begin{equation}
a_1=\frac{1}{2+q}\left[1+q-2q^2-\frac{(1+q)^{\frac{2}{3}}}{(q^4-q^3-3q^2-4q-2)^{\frac{1}{6}}}\right]\,,\qquad q< 1-\sqrt{3} \quad \mbox{or}\quad q> 1+\sqrt{3}
\end{equation}
and
\begin{equation}
a_2=\frac{1}{2-q}\left(\frac{1+q}{q^4-q^3-3q^2-4q-2}\right)^{\frac{1}{3}}\,,
\end{equation}
and from Eq. (\ref{VL}),
\begin{equation}
V(r)=-2(q^4-q^3-3q^2-4q-2)r^{\frac{2}{2+q}-2a_1}\,B(r)\,.
\end{equation}
This is in fact the class of Liefshitz solutions discussed in Ref. \cite{Weyl}.
To explicitly reconstruct the model $F=F(R)$ which generates these solutions, we use Eq. (\ref{R}) to find $r$ as a function of $R$ and one of the EoMs (\ref{uno0}). For example, for $q=4$, one has that  $F(R)\propto\sqrt{k/R}$ generates the SSS solution $B(r)=-k/7+C_0/r^2+C_1/r^7$ and $V(r)=V_0 (r/\tilde r)^7B(r)$, $V_0$ being a constant.

Despite the fact that following this procedure we can generate a large number of vacuum $F(R)$--SSS solutions, the possible choices of $F_R$ are limited by those simple cases where Eqs. (\ref{Seba1}) are not transcendental. In particular, this procedure is useful to describe black hole solutions with $V(r)=\mathrm{e}^{\alpha(r)}B(r)$, $\alpha(r)$ being a suitable function of $r$.

In the next Section, we will introduce the metric form for traversable wormholes.

\section{Traversable wormhole parametrization}

To describe a time independent, non rotating, traversable wormhole, we introduce the line element
\begin{equation}
ds^2 = - \mathrm{e}^{\phi(l)} dt^2 + dl^2 + r^2(l) d\sigma_{k}^2\label{M2}
\end{equation}
where  $l \in (-\infty,+\infty)$ and $r(l)$ is supposed to have at least one minimum $r_0$ which, without loss of generality, can occur at $l=0$. In order to avoid event horizons of sort, we shall assume $\phi(l)$ to be finite everywhere and metric components at least twice differentiable with respect to  $l$.\\
These are merely the minimal requirements to obtain a wormhole that is ``traversable in principle''. By this expression, we mean that we look for wormhole solutions with no event horizons and not based on naked singularities. It is understood that for ``realistic traversable'' models one should add technical features we are not going to discuss at this point.\\
It is worth to notice that reference to the asymptotic behavior of the solution is not addressed here. 
In General Relativity one is typically concerned with asymptotic flatness where instead $F(R)$--solutions hardly meet this requirement. 
Many viable $F(R)$-gravity models representing a realistic scenario to account for dark energy
have been proposed in the last years. These models must satisfy a list of viability conditions 
(positive definiteness of the effective gravitational coupling, matter stability condition, Solar-system constraints etc.).
Some simple examples of $F(R)$-viable models can be found in Ref.~\cite{viablemodels}, where a correction term to the Hilbert-Einstein action is added as $F (R)=R+f (R)$, being $f (R)$ a generic function of the Ricci scalar which plays the role of an effective cosmological constant.
%For such a kind of models which explicitly mimic the $\Lambda$CDM model, we do not expect %to find wormhole solutions different to the ones of General Relativity supported by matter. 
However, since the interest in modified gravity is not limited to the possibility of reproducing the dark energy epoch, but we may find many other applications (for example, related to inflation, quantum corrections in the early stage of the universe, string-inspired gravities, ...),   
we will carried out our analysis without any particular restriction on the feature of the models out of the wormhole solutions.
\\

\noindent The Ricci scalar corresponding to (\ref{M2}) reads
\begin{eqnarray}
\label{ricci2}
R&=&-\frac{1}{2r(l)^2}\left(2\phi''(l)r(l)^2+\phi'(l)^2r(l)^2
+4\phi'(l)r'(l)r(l)+8r(l) r''(l)\right.\nonumber\\
&&\left.+4r'(l)^2-4k\right)\,,
\end{eqnarray}
where, the prime $^{\prime}$ means derivative with respect to $l$. The EoMs become
\begin{equation}
\left\{\begin{array}{ll}
0= & r^2 (F-F_R R) + 2 F_R \left(k - r'^2 - 2 r r''\right) -  4 rr'\frac{dF_R}{dl}- 2 r^2  \frac{d^2 F_R}{dl^2}  \\
\\
0= &  r^2 (F-F_R R) -r F_R\left(2 r'' + r \phi'' + r' \phi' + \frac{1}{2} r \phi'^2   \right) +\\ 
\\
& \hspace{3cm}- r  \frac{dF_R}{dl}\left(2 r' + r \phi'\right)- 2 r^2 \frac{d^2 F_R}{dl^2}\,.
\end{array}\right.
\label{rdc2}
\end{equation}
%Note that with respect to the set of EoMs (\ref{uno0})--(\ref{due0}), the EOMs (\ref{rdc1}) and %(\ref{rdc2}) do not depend on $k$. 
\medskip

An alternative parametrization of the wormhole in terms of the line element (\ref{M1}) can be obtained by replacing $V(r)$ and $B(r)$ with four new functions $\phi_\pm (r)$ and $b_\pm(r)$, according to
\begin{equation}
\left\{\begin{array}{c}
V(r) = \exp( 2\phi_\pm(r)),  \qquad r \geq r_0\\
\\
B(r) = 1-\frac{b_\pm(r)}{r}, \qquad r \geq r_0\,.
\end{array}\right.
\label{b}
\end{equation}
This corresponds to the fact that the two pairs $(\phi_\pm(r),b_\pm(r))$ actually describe two different universes joined together at the throat of the wormhole, $r_0$. \\
In this case, the Ricci scalar is
\begin{equation}
R=\frac{1}{r^2}\left[
2(k-1)+(3b_\pm(r)-4r)\phi_\pm'(r)+b_\pm'(r)(2+r\phi'_\pm(r))-2r(r-b_\pm(r))(\phi_\pm'(r)^2+\phi_\pm''(r))
\right]\,,
\end{equation}
and the EoMs become
\begin{equation}
\left\{\begin{array}{ll}
0&=r^2 (F-F_R R) + 2 [k-1 + b'_\pm(r)] F_R - [4  r - 3 b_\pm(r) - r b'_\pm(r)]\frac{d F_R}{dr} + \\
\\
&- 2 r (  r - b_\pm(r)) \frac{ d^2 F_R}{dr^2}\,,\\
\phantom{line} \\
0&=r^2 (F-F_R R) + 2 \left[k-1 +\frac{ b_\pm(r)}{r} - 2 \phi'_\pm(r) (  r-b_\pm(r))\right]F_R+\\
\\
& - 2[  r-b_\pm(r)][2 + r \phi'_\pm(r)] \frac{d F_R}{dr}\,,
\end{array}\right.
\label{uno}
\end{equation}
with the requirements that \cite{visser}
\begin{enumerate}
\item $\phi_\pm(r)$, $b_\pm(r)$ are well defined for all $r \geq r_0$
\item $\phi'_+(r_0) =\phi'_-(r_0)\equiv\phi'(r_0)$
\item $b_\pm(r_0)=  r_0$ and $b_\pm(r)<  r$ for all $r>r_0$
\item $b'_+(r_0)= b'_-(r_0) <1$ 
\end{enumerate}
Although written in a different fashion, these requirements simply reproduce the physical conditions of a traversable wormhole given after equation (\ref{M2}). For sake of simplicity, one could add also the condition that the time coordinate is continuous across the wormhole throat, i.e. $\phi_+(r_0) =\phi_-(r_0)\equiv\phi(r_0)$.

%Metrics coordinates in (\ref{M1}) and (\ref{M2}) are related by 
%\begin{equation}
%V(r)=\mathrm{e}^{\phi(l)}\,,\quad B(r)=\frac{ dr(l)}{d l}\,.
%\end{equation}

\section{Effective energy conditions}

It is well known that in Einstein's gravity (namely, $F(R)=R$) vacuum wormhole solutions do not exist and the weak energy condition (WEC) must be violated by static wormholes near to the throat~\cite{mt,GR}. It means that, by introducing the stress energy tensor of matter $T_{\mu\nu}$ such that $T_\mu^\nu=\text{diag}(-\rho,p^r,p^\theta,p^\varphi)$, $\rho$ and $p^{r,\theta,\varphi}$ being the energy density and the pressure components of matter, respectively, the following relation is violated near to the throat
\begin{equation}
T_{\mu\nu}u^\mu u^{\nu}\ge0\,,\quad u^\mu \quad\text{time-like vector ($u^\mu u_\mu=-1$)}\,,\nonumber
\end{equation}
namely
\begin{equation}
\rho\ge 0\,,\quad\rho+p^{i}\ge 0\, \qquad \forall i= r,\theta,\varphi \;.\label{SEC}
\end{equation}
In a different theory of gravity, the region around the throat may respect this condition, as it has been explicitly demonstrated in Ref.~\cite{Furey:2004rq}, where a wormhole solution supported by matter which satisfies the WEC has been presented. In such a case, the geometry plays the role of repulsive gravity necessary to construct the traversable wormhole. 
In principle, any modification to Einstein's gravity of the type under investigation (\ref{action}) may be considered as an effective fluid (see also \cite{LL} for the coupling of General Relativity with a nonlinear fluid), and in the case of vacuum solutions we expect that such effective fluid violates the WEC near to the throat.

The field equations of $F(R)$-modified gravity theories may be rewritten in the form 
\begin{equation}
G_{\mu\nu}=T^{\mathrm{(MG)}}_{\mu\nu}\,,
\end{equation}
where $G_{\mu\nu}$ is the Einstein tensor
and $T^{\mathrm{(MG)}}_{\mu\nu}$ is a suitable `modified gravity'  tensor which encodes the gravity modification. Since for the metric (\ref{M1}) with relations (\ref{b}) the null zero-components of the Einstein tensor are 
\begin{eqnarray}
G_{00}&=&\frac{\mathrm{e}^{2\phi_{\pm}(r)}}{r^2}(k-1+b'_{\pm}(r))\,,\nonumber\\
G_{11}&=&-\frac{1}{(r-b_{\pm}(r))r}\left[(k-1)+\frac{ b_\pm(r)}{r} - 2 \phi'_\pm(r) (r-b_\pm(r))\right]\,,\nonumber\\
G_{22}&=&\frac{1}{4}\frac{1}{(1-k\varrho^2)}\left[\frac{4}{r}(r-b_{\pm}(r))r\phi_{\pm}'(r)+2\left(\frac{b(r)}{r}-b'(r)\right)\right.\nonumber\\&&\left.
+r(r-b_{\pm})(4\phi_{\pm}'(r)^2+4\phi_{\pm}''(r))
+2\phi_{\pm}'(r)(b_{\pm}(r)-b_{\pm}'(r)r)\phantom{\frac{0}{0}}\right]\nonumber\,,\\
G_{33}&=&(1-k\varrho^2)\varrho^2G_{22}\,,
\end{eqnarray}
one has that the EoMs (\ref{uno}) correspond to
\begin{eqnarray}
T^{\mathrm{(MG)}}_{00}&=&-\frac{\mathrm{e}^{2\phi_{\pm}(r)}}{2r^2}\left(r^2 (F-F_R R) + 2 (F_R-1) [(k-1)+b'_\pm(r)]
\right.\nonumber\\
&&\left.- F_R'[4 r - 3 b_\pm(r) - r b'_\pm(r)]-2 r (r - b_\pm(r))  F''_R\right)\,,\nonumber\\\nonumber\\
T^{\mathrm{(MG)}}_{11}&=&\frac{1}{2(r-b_{\pm}(r))r}\left(
r^2 (F-F_R R) + 2(F_R-1)\left[(k-1)\right.\right.\nonumber\\
&&\hspace{-2cm}\left.\left.+\frac{ b_\pm(r)}{r} - 2 \phi'_\pm(r) (r-b_\pm(r))\right]\phantom{\frac{0}{0}}- 2(r-b_\pm(r))(2 + r \phi'_\pm(r)) F'_R\right)\,,
\end{eqnarray}
and
\begin{eqnarray}
T_{22}^{\mathrm{(MG)}}&=&\frac{1}{4(1-k\varrho^2)}\left[\phantom{\frac{0}{0}}2r^2(F-(F_R-1)R)-4(1-F_R)\times\right.
\nonumber\\
&&\times\left(k-1+\frac{b_{\pm}}{2r}+\frac{b'_{\pm}(r)}{2}-\phi_{\pm}'(r)(r-b_{\pm}(r))\right)+
\nonumber\\
&&\hspace{-2cm}\left.-F_R'\left(b_{\pm}(r)-rb'_{\pm}(r)-2\phi_{\pm}'(r)r^2+2\phi_{\pm}'(r)rb_{\pm}(r)\right)
-2rF''_R(r-b_{\pm}(r))\phantom{\frac{0}{0}}\right]\,,\nonumber\\
T_{33}^{\mathrm{(MG)}}&=&(1-k\varrho^2)\varrho^2T_{22}^{\mathrm{(MG)}}\,,
\end{eqnarray}
as a consequence of the summation of the two EoMs.
We can now define an effective energy density $\rho_{\mathrm{eff}}$ and effective pressures $p_{\mathrm{eff}}^{r,\varrho,\varphi}$ given by modified gravity in the form  $T^{\mathrm{(MG)}\nu}_{\mu}=\mathrm{diag}(-\rho_{\mathrm{eff}}, p_{\mathrm{eff}}^r, p_{\mathrm{eff}}^\varrho, p_{\mathrm{eff}}^\varphi)$.
%\begin{equation}
%\rho_{\mathrm{eff}}:=\frac{T_{00}}{g_{00}}\equiv\mathrm{e}^{-2\phi(r)}T_{00}\quad
%p_{\mathrm{eff}}:=\frac{T_{11}}{g_{11}}\equiv \frac{(r-b_{\pm})}{r}T_{00}\,.\label{relations}
%\end{equation}

Let us suppose to have found a wormhole solution 
characterized by $b(r)$ and $\phi(r)$.
At the throat $r_0$, such that $b(r_0)=r_0$, from the EoMs we get
\begin{eqnarray}
F_0&=&R_0 F_{R_0}-\frac{2F_{R_0}k}{r_0^2}\,,\\
F'_{R_0}&=&-\frac{2F_{R_0}}{r_0}\,.
\end{eqnarray}
We use the suffix `0' for all quantities evaluated on the throat.
Then, by taking the derivatives of the EoMs, we also have
\begin{eqnarray}
R_0&=&\frac{4k}{r_0^2}-\frac{3}{r_0^2}+\frac{3b'(r_0)}{r_0^2}\,,\label{R2}\\ 
R_0&=&\frac{4k}{r_0^2}+\frac{6}{r_0^2}\left(1-b'(r_0)\right)+3\left(1-b'(r_0)\right)\frac{F''_{R_0}}{2F_{R_0}}\,,
\end{eqnarray}
and, as a consequence,
\begin{equation}
F_{R_{0}}''=\frac{2F_{R_0}}{r_0^2}\,.
\end{equation}
By combining Eq. (\ref{R}) and Eq. (\ref{R2}) we get
\begin{equation}
\phi'(r_0)=\left(\frac{-b'(r_0)+1-2k}{r_0-b'(r_0)r_0}
\right)\,.
\end{equation}
From above relations (in particular, by using Eq. (\ref{R2})) we get
\begin{eqnarray}
\rho_{\mathrm{eff}}&:=&-\frac{T_{00}}{g_{00}}\equiv\frac{r_0^2 R_0-k}{3r_0^2}\,,\nonumber\\
p_{\mathrm{eff}}^r&:=&\frac{T_{11}}{g_{11}}\equiv-\frac{k}{r_0^2}\,,\nonumber\\
p_{\mathrm{eff}}^{\varrho,\varphi}&:=&\frac{T_{22}}{g_{22}}=\frac{T_{33}}{g_{33}}\equiv\frac{k-r_0^2R_0}{3r_0^2}\,,
\end{eqnarray}
according with Ref. \cite{cond} in the topological case $k=1$. Now, by using (\ref{R2}), we have
\begin{equation}
\rho_{\mathrm{eff}}+p_{\mathrm{eff}}^{r}=\frac{3}{r_0^2}\left[b'(r_0)-1\right]\,,
\end{equation}
and, due to the fact that for traversable wormholes $b'(r_0)<1$, the WEC (\ref{SEC}) is violated.

Some important remarks on energy conditions in $F(R)$-gravity theories can be found in Ref.~\cite{F(R)energy}.

\section{Solutions with constant curvature}

In this Section, at first we will consider the simple case of solutions with $R=0$ with the possibility to reproduce vacuum wormhole solutions. Then, a generalization to the case of constant Ricci scalar different to zero will be investigated. 

If $F(R)=R \,g(R)$ such that $\lim_{R\rightarrow 0}g(R)=0$,
it is easy to see that the EoMs trivially are satisfied for $R=0$.  
We remark that this  {\it ansatz} includes solutions of a a large class of interesting $F(R)$--gravity models, in particular Lagrangian of the type $F(R)\propto R^n\,,n\geq 2$. 
This kind of terms based on the power law of the Ricci scalar has been often studied in literature (some important features are related to the possibility to support the early time inflation, to protect the theory against divergences and singularities...).
In addition, this form of the Lagrangian possesses the Schwarzshild solution and the important class of black hole Clifton-Barrow solutions.
In some sense  we can say that these solutions are trivial in the measure that is the Schwarzschild black hole.

\paragraph{Solution n.1:}

\begin{equation}
\left\{\begin{array}{ll} 
V(r) &\equiv \mathrm{exp}(2\phi_\pm(r))= \left(\frac{r}{\tilde r}\right)^q\,,\\
\\
b_\pm(r)&=\frac{kq(2+q)r}{4+ 2q + q^2}+ C_0\, r^{-\frac{q(1+q)}{4+q}}\,,
\end{array} \right.
\label{wh1}
\end{equation}
where $\tilde r$ is a scale, $q$ a dimensionless parameter and $C_0$ a positive integration constant. It is not difficult to show that requirements from 1. to 4. above are fulfilled by this solution for any $k$ if and only if $q>-4$. If this is the case, the wormhole throat is located at
$$
r_0 = \left[ \frac{q^2+2q+4}{(1-k) q^2 + 2 (1-k)q  +4}\,C_0\right]^{\frac{q+4}{q^2+2q+4}}\;.
$$

\paragraph{Solution n.2:}
 \begin{equation}
\left\{\begin{array}{ll} 
V(r)&\equiv \mathrm{exp}(2\phi_\pm(r)) =\mathrm{e}^{\frac{2\tilde r}{r}}\,,\\
b_\pm(r)&=\frac{2 k \tilde r^2}{\left(2-\frac{\tilde r}{r}\right)^5}\left(\frac{71}{2r}-\frac{103}{4\tilde r}-\frac{39\tilde r}{2r^2}+\frac{5\tilde r^2}{r^3}-\frac{\tilde r^3}{2r^4}\right)+\frac{C_1}{\left(2-\frac{\tilde r}{r}\right)^5}\,\mathrm{e}^{-\frac{2\tilde r}{r}}\,, 
\label{wh2}
\end{array}\right.
\end{equation}
where $\tilde r$ is a dimensional parameter and $C_1$ the integration constant of the solution.

\paragraph{Solution n.3:} only if $k=1$, referring to the metric form (\ref{M2}),
\begin{equation}
\label{wh3_0}
\left\{\begin{array}{ll} 
r(l)&=\sqrt{r_0^2+l^2} \\
\mathrm{e}^{\phi(l)}&=\Phi_0 \left(\frac{r_0}{r(l)}\right)^2
\end{array}\right.
\end{equation}
where $r_0$ explicitly represents the wormhole throat,  $\Phi_0 >0 $ is an opportune dimensionless constant of integration which basically determine the flow of time at the throat $l \rightarrow 0$ and at infinity $l\rightarrow \pm\infty $.\\
\\
Let us consider now a generalization of solution (\ref{wh3_0}) to the case of constant $R=R_0$, but with $R\neq 0$. 

\paragraph{Solution n.4:} By putting $R=R_0$ in (\ref{ricci2}), it is easy to see that for the topological case $k=1$ one possibility is given by
\begin{equation}
\label{wh3_1}
\left\{\begin{array}{ll} 
r(l)&=\sqrt{r_0^2+l^2} \\
\mathrm{e}^{\phi(l)}&=\Phi_0 \left(\frac{r_0}{r(l)}\right)^2\cos\left(\frac{l\sqrt{R_0}}{2}+\Phi_1\right)\,,
\end{array}\right.
\end{equation}
where $\Phi_{0,1}$ are constants and for $R_0=0$ one recovers solution (\ref{wh3_0}). Here, one important remark is in order. Since the metric parameter $\exp\left[\phi(l)\right]$ must be different to zero for any point of the space time, we must require $R_0<0$, such that
\begin{equation}
\mathrm{e}^{\phi(l)}=\Phi_0 \left(\frac{r_0}{r(l)}\right)^2\cosh\left(\frac{l\sqrt{|R_0|}}{2}+\Phi_1\right)\,,\quad R_0<0\,.
\end{equation}
It represents a wormhole solution with constant and negative curvature $R_0$. Lagrangians of the type $F(R)\propto (R-R_0)^n$, $n\geq 2$ possess this kind of solution (in this case, $F(R_0)=F'(R_0)=0$ and the EOMs are trivially satisfied). It is interesting to note that this models
also have the topological Schwarzschild-de Sitter solution (\ref{M1}) with $V(r)=B(r)=k-C/r-\Lambda r^2/3$, where $C$ is an integration constant and $\Lambda$ is given by $\Lambda=R_0/(4-2n)>0$, since in this case the two EOMs (\ref{uno0}) are equal and satisfied (remember that $R=4\Lambda$ on Schwarzschild-de Sitter solution).

\section{Solutions with non constant curvature}

In this Section, we will furnish the formalism and the implicit form of the $F(R)$-gravity models where vacuum wormhole solutions with non constant curvature can be realized. Despite to the fact that the explicit forms of the models are hard to be derived, they may be accessible via numerical analysis.
\medskip

With reference to equations (\ref{M2}), (\ref{ricci2}) and (\ref{rdc2}), let us re-write the EoMs more explicitly, as:
\begin{equation}
\left\{\begin{array}{ll}
0= & F + F_R \left(\phi'' + \frac{1}{2} \phi'^2 + \frac{2 \phi' r'}{r}\right) -  \frac{4r'}{r} \frac{dF_R}{dl}- 2  \frac{d^2 F_R}{dl^2}  \\
\\
0= &  F- F_R \left(\frac{2 k}{r^2} - \frac{2 r'^2}{r^2} - \frac{2 r''}{r} -\frac{ \phi' r'}{r}\right) - \left(\phi' + \frac{2r'}{r}\right)  \frac{dF_R}{dl}- 2 \frac{d^2 F_R}{dl^2}\,,
\end{array}\right.
\label{rdc3}
\end{equation}
where it is understood that $\phi$ and $r$ are functions of $l$. Subtracting one from the other and integrating, we get\\
\phantom{line}
 \begin{equation}
\frac{F_R(l)}{^{0}F_R } = \exp \int^l d\tilde l \,\frac{\phi'' +  \frac{1}{2}\phi'^2 +  \phi'r'/r + 2 k/r^2- 2r'^2/r^2 - 2 r''/r}{2r'/r -\phi'} \equiv \exp \int^l  \pi(\tilde l) d\tilde l\;,
\label{pi}
 \end{equation}
\phantom{line}\\
 $^{0}F_R$ being an integration constant. Summing up the two EoMs, we have\\
\phantom{line}
 \begin{equation}
0= 2F + F_R \left(\phi''+\frac{1}{2}\phi'^2- \frac{3 \phi' r'}{r} - \frac{2k}{r^2} + \frac{2r'^2}{r^2} + \frac{2 r''}{r} - \phi' \pi - \frac{6\pi r'}{r} - 4 \pi^2 + 4\pi'  \right)\,,
\end{equation}
\phantom{line}\\
that is
\begin{equation}
 F = F_R \Delta(l),
\end{equation}
expression that can be assumed as the implicit definition of $\Delta (l)$. On the other hand,
\begin{equation}
F_R = \frac{dF}{dR} = \left(\frac{1}{R'(l)} \right)\frac{d F(l)}{dl}\,,\label{43}
\end{equation}
and then
\begin{equation}
\frac{F(l)}{^{0}F}\equiv \exp\int^l \, d\tilde l\;\frac{R'(\tilde l)}{\Delta(\tilde l)} \label{delta}
\end{equation}
where $^{0}F$ is an integration constant. Combining (\ref{pi}), (\ref{43}) and (\ref{delta}), we find that 
\begin{equation}
R'(l) = \Delta'(l) + \pi(l) \Delta(l)\,. \label{vie}
\end{equation}
Note that $[\pi(l)]=[1/l]$ and $[\Delta(l)]=[1/l^2]$. Thus, because of (\ref{delta}) and (\ref{vie}), one has
\begin{equation}
F(l) =\left( \frac{^{0}F}{^{0}\Delta}\right) \Delta(l) \exp \int^l d\tilde l\, \pi(\tilde l)\,,\label{last}
\end{equation}
and also $^{0}\Delta$ is an integration constant. 
It is possible to verify that, by using (\ref{vie}), this expression is consistent with (\ref{pi}) provided that the integration constants $^{0}F_R$, $^{0}F/^{0}\Delta$ be equal. We stress that
up to this point the description is completely model independent.

To describe a wormhole, we need to consider some well-defined, twice differentiable function $r(l)$ with at least one minimum. Mimicking the General relativity standard wormhole, we could think to look for a similar function among the two-parameter family 
\begin{equation}
r(l) := \left(l^{2n} + r_0^{2n}\right)^{\frac{1}{2m}}\label{radius}\,,
\end{equation}
with $n$ integer and greater than 1 and a $m$ positive rational number satisfying the EoMs (we recover (\ref{wh3_0}) for $n=1$). This solution would represent a wormhole with the throat placed at $l=0$ and an asymptotic behavior strongly dependent on the values of $(n,m)$ but presumably far from asymptotic flatness.
It turns out that 
\begin{equation}
\label{deriv}
r'(l) =\left(\frac{n}{m}\right)\frac{l^{2n-1}}{ r(l)^{2m-1}}\,, \qquad r''(l) = \left(\frac{n}{m}\right)\frac{l^{2(n-1)}}{r^{2m-1}(l)}\left[\frac{n}{m} -1 + (2n-1)\frac{r_0^{2n}}{r^{2m}(l)}\right]\,.
\end{equation}
The scalar curvature (\ref{ricci2}) reads
\begin{eqnarray}
R(l) &=&\frac{r(l)^{-2(2m+1)}}{4l^2 m^2}\left[-4l^{2n+1} m\,n\, r(l)^{2(m+1)}\phi'(l) 
-l^2m^2 r(l)^{2(2m+1)}\phi'(l)^2+
4kl^2 m^2 r(l)^{4m}
\right.\nonumber\\&&\left.
-4l^{2n}n\,r(l)^{2}\left(l^{2n}(3n-2m)+2m(2n-1)r_0^{2n}\right)-2l^2m^2 r(l)^{2(2m+1)}\phi''(l)\right]\,.
\end{eqnarray}
Then, for $\pi(l)$, $\Delta(l)$ one derives
\begin{eqnarray}
\pi(l) &=& 
\frac{
\frac{n\phi'}{l\,m(1+l^{-2n}r_0^{2n})}
+\frac{r^{-2(2m+1)}}{l^2m^2}\left(2k\,l^2m^2r^{4m}-2l^{2n}n\,r^2(l^{2n}(2n-m)+2m(2n-1)r_0^{2n}\right)}
{\frac{2n}{l\,m(1+l^{-2n}r_0^{2n})}-\phi'}\nonumber\\
&&
+\frac{\frac{\phi'}{2}+l^2m^2r^{2(2m+1)}\phi''}{\frac{2n}{l\,m(1+l^{-2n}r_0^{2n})}-\phi'}\,,\nonumber\\
\Delta(l)&=&
-\frac{2l^{2(2n-1)}n^2}{m^2\,r^{4m}}+\frac{2l^{2(2n-1)n^2}}{m\,r^{4m}}+\frac{l^{2(n-1)}}{m\,r^{2m}}-\frac{2l^{2(n-1)}n^2}{m\,r^{2m}}+\frac{k}{r^2}-\frac{\phi'^2}{4}-\frac{\phi''}{2}\nonumber\\&&
+\frac{3n\phi'}{2lm(1+l^{-2n}r_0^{2n})}-2\pi'(l)^2+2\pi(l)^2+
\frac{\pi(l)}{2}\left(\frac{6n}{l\,m(1+l^{-2n}r_0^{2n})}+\phi'\right)\,,
\end{eqnarray}
and in principle one can solve Eq.~(\ref{vie}). 

Let us summarize the result. The topological wormhole solutions (\ref{M2}) in empty space-time  with $r(l)$ in the (appropriate) form of (\ref{radius}) can be realized in $F(R)$-gravity consistently with Eq.~(\ref{vie}), which is in fact a differential equation for $\phi(l)$.
Thus, the implicit form of the model is given by (\ref{last}). In principle, given the Ricci scalar as a function of $l$, one can try to reconstruct the models possessing these solutions.
However, it is not possible to solve Eq.~(\ref{vie}) in an analytical way, and some specific choice of $(n,m)$ in (\ref{radius}) and of the topology must be done. Thus, by introducing some boundary conditions, numerical calculations might be implemented in this formalism.

\section{Conclusions}

In this paper, we have considered topological wormhole solutions in $F(R)$-gravity, motivated by the popularity of this kind of modifications to Einstein's gravity and by the importance to recover 
such a kind of objects in their framework. Since in General Relativity traversable wormholes have to been supported by a matter source which violates the energy conditions in a region around the throat, it may be interesting to see if empty traversable wormholes exist in modified theories of gravity, where the geometry plays the role of repulsive gravity. At this regard, we explicitly show that effective energy conditions are violated on the throat in $F(R)$-gravity, generalizing the results already present in literature to the topological case. 
The formalism of traversable wormhole metric in $F(R)$-gravity has been derived by using a method based on the Lagrangian multipliers, which permits to deal with a system of ordinary differential equations. The metric is presented in two suitable equivalent forms. 
Despite to the fact that $F(R)$-black hole solutions can be easily reconstructed by starting from the metric, vacuum wormhole solutions are much more difficult to be found, except for the case of null (or constant) Ricci scalar. For this case, we have found several solutions. We would like to stress that this kind of solutions can be realized by a large class of $F(R)$-models: for example, this is the case of Lagrangians of the type $F(R)\propto R^n\,,n\geq 2$, which are used in inflationary scenario in the attempt to reproduce the early-time acceleration, and as a consequence it may be interesting
to investigate the possibility of having wormholes in primordial universe. In the last part of the paper, we had a look to wormhole solutions with non-zero curvature and implicit form of $F(R)$-models which realize these solutions is derived. Here, exact solutions cannot be found 
in analytical way. However, numerical calculations might be implemented in the formalism, an interesting task for future works.

We end with the following consideration. $F(R)$-gravity not only represents a possible alternative to cosmological constant to explain current acceleration of the universe, but it may be understood also as an effective action coming from a still unknown quantum theory of gravity. In this sense, $F(R)$ would take into account quantum corrections to classical theory, corrections which probably should also account for wormhole production (e.g. \cite{hw}).  In this respect, the question posed in this paper is all but useless, and it might be of interest to study what is the most important 
contribution in primordial wormholes creation: whether modified gravity or quantum 
effects of  GUTs, as investigated in Ref.~\cite{concl}.

\section*{Acknowledgments}
We thank K. A. Bronnikov for valuable comments and suggestions. \\
RDC wishes to acknowledge with gratitude the financial support and friendly hospitality of the {\it Interdisciplinary Laboratory for Computational Science} (LISC), FBK--CMM, Trento -- Italy  where part of this work has been done.

\end{document}